\documentclass[conference]{IEEEtran}
\IEEEoverridecommandlockouts

\usepackage{cite}
\usepackage{amsmath,amssymb,amsfonts}
\usepackage{algorithm}
\usepackage{algorithmic}
\usepackage{graphicx}
\usepackage{textcomp}
\usepackage{xcolor}
\usepackage{url}
\def\BibTeX{{\rm B\kern-.05em{\sc i\kern-.025em b}\kern-.08em
    T\kern-.1667em\lower.7ex\hbox{E}\kern-.125emX}}
\DeclareRobustCommand*{\IEEEauthorrefmark}[1]{%
    \raisebox{0pt}[0pt][0pt]{\textsuperscript{\footnotesize\ensuremath{#1}}}}

\begin{document}

\title{Matrix Factorization with Dynamic Multi-view Clustering for Recommender System\\
}
\author{
\IEEEauthorblockN{
Shangde Gao\IEEEauthorrefmark{1,}\IEEEauthorrefmark{2,\dagger},
Ke Liu\IEEEauthorrefmark{1,\dagger},
Yichao Fu\IEEEauthorrefmark{1}, 
Hongxia Xu\IEEEauthorrefmark{2}, and
Jian Wu\IEEEauthorrefmark{3,}\IEEEauthorrefmark{4,}\IEEEauthorrefmark{5}}
\IEEEauthorblockA{\IEEEauthorrefmark{1}College of Computer Science and Technology, Zhejiang University, Hangzhou, 310058, China}
\IEEEauthorblockA{\IEEEauthorrefmark{2}Liangzhu Laboratory and WeDoctor Cloud, Hangzhou, 310058, China}
\IEEEauthorblockA{\IEEEauthorrefmark{3}The Second Affiliated Hospital, Zhejiang University School of Medicine, Hangzhou, 310058, China}
\IEEEauthorblockA{\IEEEauthorrefmark{4}State Key Laboratory of Transvascular Implantation Devices and TIDRI, Hangzhou, 310009, China}
\IEEEauthorblockA{\IEEEauthorrefmark{5}Zhejiang Key Laboratory of Medical Imaging Artificial Intelligence, Hangzhou, 310058, China}
\IEEEauthorblockA{\{gaosde, lk2017, fuyichao, Einstein, wujian2000\}@zju.edu.cn}
\thanks{$\dagger$: Equal-contribution authors.}
\thanks{Corresponding author: Hongxia Xu (Einstein@zju.edu.cn)}
}


\maketitle

\begin{abstract}
Matrix factorization (MF), a cornerstone of recommender systems, decomposes user-item interaction matrices into latent representations. Traditional MF approaches, however, employ a two-stage, non-end-to-end paradigm, sequentially performing recommendation and clustering, resulting in prohibitive computational costs for large-scale applications like e-commerce and IoT, where billions of users interact with trillions of items. To address this, we propose Matrix Factorization with Dynamic Multi-view Clustering (MFDMC), a unified framework that balances efficient end-to-end training with comprehensive utilization of web-scale data and enhances interpretability. MFDMC leverages dynamic multi-view clustering to learn user and item representations, adaptively pruning poorly formed clusters. Each entity's representation is modeled as a weighted projection of robust clusters, capturing its diverse roles across views. This design maximizes representation space utilization, improves interpretability, and ensures resilience for downstream tasks. Extensive experiments demonstrate MFDMC's superior performance in recommender systems and other representation learning domains, such as computer vision, highlighting its scalability and versatility.
\end{abstract}


\section{Introduction}
\label{sec:intro}
Recommendation systems have become increasingly essential, driven by the rapid expansion of network services such as e-commerce, social networks, and digital entertainment services~\cite{yu2023self,ko2022survey,zhao2024recommender}. 
These systems assist users in discovering items of interest from a vast array of options, improving personalization and decision-making. They also boost commercial value by increasing engagement and fostering loyalty.

Matrix factorization (MF) \cite{isinkaye2023matrix,bobadilla2024comprehensive} is a pivotal technique in recommendation systems, capable of effectively extracting latent user preferences and item features from historical data. The fundamental principle of MF is based on singular value decomposition, which maps users and items into a shared latent space and models user-item interactions through inner products. However, traditional MF methods exhibit significant limitations when addressing challenges such as high computational complexity, cold-start problems, and data sparsity.~\cite{isinkaye2023matrix, xiang2024neural,surprise,biased_mf,xu2020tt,zhu2023self,zhang2024federated}.
Specifically, FunckMF \cite{surprise} and biased MF \cite{biased_mf}, which improve \textit{rating prediction} accuracy by introducing rating bias mechanisms, incur substantial computational costs when processing large-scale user-item data, especially in scenarios involving billions of users and items, making it difficult to meet real-time recommendation demands \cite{wu2023personalized, guo2024paper}. 
As for the \textit{cold-start problem}, Asymmetric SVD \cite{xu2020tt,zhu2023self} struggle to generate accurate recommendations for new users or items because lacking of historical interaction data, their effectiveness remains limited by data sparsity and model expressiveness.

With the rapid development of DNNs~\cite{gao2023contrastive,gao2024collaborative}, MF with different deep learning methods to tackle the challenges of high computational complexity, cold-start problems, or data sparsity, has become a research topic in recommender systems \cite{zhao2017multi,li2023survey,chen2023distribution,chen2023improving,wu2024self,wang2023generalized}.
Specifically, \textit{multi-view clustering (MVC)} methods~\cite{guo2017deepfm,gunawardena2024dccnmf} provide a robust architectural solution for learning complementary and consensus information by integrating multi-source heterogeneous data, such as user behavior, social relationships, item attributes.
\textit{Graph neural network methods (GNNs)} have been proposed to exploit user-item interaction graphs for improved recommendation accuracy\cite{chen2023distribution,sun2021interest}. However, the performance degrades in data sparsity due to strong reliance on graph structures.
\textit{Privacy-preserving MF methods} leveraging federated learning frameworks have been developed to ensure data security while maintaining performance \cite{li2023survey}. Furthermore, attention-based mechanisms have been integrated into MF to capture contextual dependencies and enhance the modeling of user-item interactions \cite{wu2022attention,sun2022learning,yang2023empowering}. 

Despite these progress, existing methods still have some open challenges. For instance, while MF methods based on Deep Neural Networks (DNNs) enhance recommendation performance, their "black-box" nature leads to poor interpretability \cite{survey1, sinha2022dnn}. Moreover, the latent representation space in DNNs may be underutilized, resulting in dimensional redundancy and information redundancy.

To address these challenges, we propose a unified \textbf{M}atrix \textbf{F}actorization with \textbf{D}ynamic \textbf{M}ulti-view \textbf{C}lustering (\textbf{MFDMC}) framework. By incorporating multi-view clustering techniques, this method can more effectively capture multi-dimensional features of users and items, while reducing computational complexity and alleviating data sparsity issues. Additionally, the dynamic nature of MFDMC enables it to adapt to changing user behaviors, thereby improving recommendation accuracy and real-time performance.
The key contributions of this research are as follows:
\begin{itemize}
    \item We present a novel MF approach, MFDMC, which integrates dynamic multi-view clustering into the matrix factorization framework. This integration leverages the representation space more effectively, resulting in substantial reductions in time and computational demands.
    \item We employ visualization techniques to assess and validate the interpretability of user and item representations generated through dynamic multi-view clustering.
    \item Comprehensive experiments demonstrate the effectiveness and efficiency of MFDMC. Extra experiments demonstrate that this idea can be easily generalized to other machine learning tasks where representation learning is essential, for example, computer vision tasks.

    
\end{itemize}

\section{Related Works}
\label{sec:related}
\subsection{Matrix Factorization}
Matrix factorization (MF) techniques, rooted in Singular Value Decomposition (SVD), have been widely adopted in recommendation systems. However, traditional SVD is inherently limited to decomposing dense matrices, while user-item interaction matrices in real-world scenarios are typically highly sparse. To address this, missing values in the rating matrix must be imputed before applying SVD. 
This approach introduces two significant challenges: (1) the imputation process can substantially increase the data volume and computational complexity, and (2) inappropriate imputation methods may lead to data distortion, adversely affecting the model's performance.
To circumvent these limitations, researchers have explored alternative matrix factorization methods that focus solely on the observed ratings, avoiding the need for imputation. Notable advancements in this direction include FunkMF \cite{web:mf}, Probabilistic Matrix Factorization (PMF) \cite{pmf}, and BiasedMF \cite{biased_mf}. FunkMF, for instance, employs stochastic gradient descent to optimize the following objective function:

\begin{equation}
    \begin{aligned}
    \label{equ:funkmf}
    \min_{p,q} \quad \sum_{(u,i)\in \mathcal{K}}(r_{ui}-q_i^{T}p_u)^2 + \lambda(\Vert q_i \Vert ^2 + \Vert p_u \Vert ^2)
    \end{aligned}
\end{equation}

Here, \( p_u \) and \( q_i \) represent the latent factors for users and items, respectively, and \( \lambda \) is a regularization parameter to prevent overfitting. BiasedMF \cite{biased_mf} extends this framework by incorporating user and item biases, as well as the global average rating, to account for systematic variations in the observed ratings. This decomposition allows the model to capture both the individual biases and the interactions between users and items more effectively.

Recent advancements have integrated deep learning techniques into matrix factorization frameworks, aiming to enhance the representational capacity of the models. For instance, studies such as \cite{wang2023generalized} and \cite{zhang2023lightfr} have demonstrated that incorporating higher-order feature interactions and non-linear relationships can significantly improve the performance and generalization capabilities of recommendation systems.
\subsection{Multi-view Learning}
Multi-view learning has been widely adopted in machine learning, enabling models to leverage diverse data representations for enhanced performance~\cite{sun2024robust,yuan2025prototype}. Notable applications include multi-head attention in Transformers \cite{10088164,zhao2017multi,li2023better,li2024fewvs}, which captures information from different subspaces simultaneously, and the Multi-Interest Network (MIP) \cite{shi2023everyone,li2023survey}, which generates multiple interest embeddings for users based on sequential interactions, learning weights to represent preference over each embedding.
Recent advancements in multi-view clustering further extend this paradigm, particularly in dynamic environments. Deep multi-view clustering method integrate feature learning and clustering to achieve unified representations across views. Additionally, \cite{wang2023generalized,gunawardena2024dccnmf} introduced dynamic multi-view clustering, adapting to evolving data distributions for real-world applications. These methods demonstrate the potential of multi-view clustering to handle non-stationary data and improve representation learning.
In this work, we extend multi-view learning to dynamic multi-view clustering for recommendation systems. Our framework captures the multiple roles of users and items by clustering their representations across views, enriching the representation space for downstream tasks. 
\subsection{Interpretability in Recommender Systems}
\begin{table}[t!]
    \centering
    \caption{Notations and descriptions}
    \begin{tabular}{c|p{0.31\textwidth}}
         \hline
         Notations & Notion \\
         \hline
         $d$ & Dimension of latent vector, $d \in \mathbb{Z}$\\
         $m, n$ & Number of users/items, $m, n \in \mathbb{Z}$\\
         $P, Q$ & Matrix of users/items, $P, Q \in \mathbb{R}^{m\times d}$\\
         $p_i, q_i$ & The $i^{th}$ user/item latent vector, $p_i, q_i\in \mathbb{R}^d$\\
         $R$ & User-item interaction matrix, $R \in \mathbb{R}^{m \times n} $\\
         $t,v$ & Number of centers, views, $e, v \in \mathbb{Z}$\\
         $b$ & Dimension of centers, $b \in \mathbb{Z}$ and $b = d/v$ \\
         $C^{user}, C^{item}$ & The centers of user/item, $C \in \mathbb{R}^{v \times e \times b}$ \\
         $c_{i,j}^{user}, c_{i,j}^{item}$ & The $i^{th}$ center in $j^{th}$ view of user/item, $c_{i,j} \in \mathbb{R}^{b}$\\
         $W^{user}_i, W^{item}_i$ & Weight of $i^{th}$ user/item for centers, $W \in \mathbb{R}^{v\times e}$\\
         $w_{i,j}^{user}, w_{i,j}^{item}$ & Weight of user/item $i^{th}$ center in $j^{th}$ view, $w_{i,j} \in \mathbb{R}$\\
         $\rho, \eta, \gamma, \psi, \lambda$ & Weight parameters\\
         \hline
    \end{tabular}
    \label{tab:my_label}
\end{table}
Few research has focused on improving the interpretability of matrix factorization (MF)-based recommendation models. Interpretability is crucial for understanding the rationale behind recommendations, particularly in domains where explainability is as important as accuracy. Zhang et al. \cite{emf} proposed an explicit factor model that extracts features from user reviews and maps each latent dimension to a specific feature, enabling dimension-level interpretability and traceability in the recommendation process. Similarly, Chen et al. \cite{explanation1} introduced a feature cube framework, which ranks features and items using a pairwise learning-to-rank method, providing insights into the factors influencing recommendations.
Further advancements include SULM \cite{expl2}, which integrates user sentiments about item features into MF to predict ratings and generate feature-level explanations. Explainable Matrix Factorization (EMF) \cite{expl3} enhances interpretability by adding an "interpretable regularizer" to the MF objective function, producing user-relevant explanations alongside recommendations. These methods highlight the importance of feature-level and sentiment-based interpretability in MF models.
Recent studies have leveraged visualization tools like SHAP (SHapley Additive exPlanations) and Saliency maps to provide post-hoc interpretability \cite{rotem2024visuala,rotem2024visualb}.
They quantify the contribution of each feature to down-tream tasks, offering intuitive visualizations of the model's decision-making process.
Inspired by this, we visualize each view of features for exploring the interpretability of our method in this paper.

\section{Method}
\begin{figure}[t!]
\centering
\includegraphics[width=0.9\linewidth]{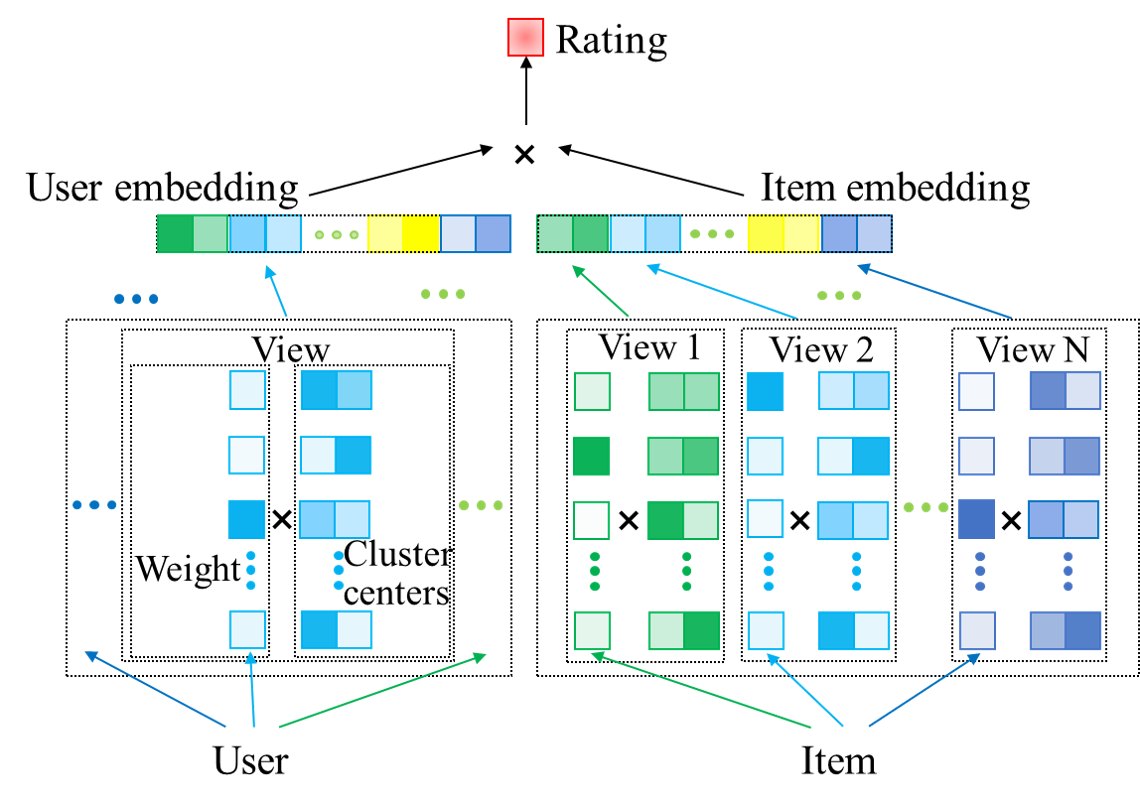}
    \caption{Schematic Illustration of the Multi-View Feature-Weighted Dynamic Cluster (MFDMC) Framework. User/item embeddings are projected into $t$ weighted cluster centers across $v$ distinct views, with each view’s representation formulated as a weighted summation of cluster centers. Distinct color groups denote different views, where darker shading intensity corresponds to higher feature importance.
    }\label{fig:model}
    \vspace*{-0.15in}
\end{figure}
In this section, we propose a Unified Matrix Factorization with Dynamic Multi-View Clustering (MFDMC) framework, which operates in an end-to-end training paradigm. The goal is to obtain \textbf{interpretable representations} of users and items while fully utilizing the representation space. The key idea is to decompose the latent representations of users and items into multiple \textbf{views of interest}, where each view corresponds to a specific aspect of user preferences or item characteristics (e.g., genre preferences in movies). The frequently used notations are summarized in Table \ref{tab:my_label}.
\subsection{Overview of MFDMC}
The architecture of MFDMC is illustrated in Fig. \ref{fig:model}. 
The user/item latent representations are divided into sub-vectors across $v$ views of interest. In each view, users and items are assigned to one of several clusters, where each cluster represents a specific preference or characteristic. The cluster centers are dynamically updated during training to ensure optimal utilization of the latent space. The latent vectors for users and items are derived as follows:
\begin{equation}
\begin{aligned}
\label{equ:sub1}
p_i = \bigoplus_{j=0}^{v} \sum_{i=0}^t c_{i,j}^{user} w_{i,j}^{user},
q_i = \bigoplus_{j=0}^{v} \sum_{i=0}^t c_{i,j}^{item} w_{i,j}^{item}
\end{aligned}\end{equation}
Here, $p_i$ and $q_i$ represent the latent vectors for users and items, respectively. $\bigoplus$ denotes the sum of vectors.
The number of views $v$ is the same for both users and items, but the number of cluster centers $e$ can vary across views. 

\subsection{Cluster centers}
The clustering mechanism in MFDMC addresses two key challenges:
(1) Full utilization of the latent space: Cluster centers should be spread out to avoid redundancy and ensure diversity.
(2) Dynamic management of cluster centers: The number of clusters should adapt during training to avoid overfitting or underfitting.
To ensure that cluster centers are well-distributed in the latent space, we introduce a spread loss that penalizes cluster centers that are too close to each other. The loss function is defined as:
\begin{equation}
\begin{aligned}
\label{equ:loss1u}
loss_1^{user} =  \sum_{j=0}^{v} \sum_{\alpha,\beta}\mathit{l}(c_{\alpha,j}^{user}, c_{\beta,j}^{user})\\
loss_1^{item} = \sum_{j=0}^{v} \sum_{\alpha,\beta}\mathit{l}(c_{\alpha,j}^{item}, c_{\beta,j}^{item})\\
\mathit{l}(c_{\alpha}, c_{\beta}) = max\{0, \rho - \mathcal{D}(c_{\alpha} - c_{\beta})\}
\end{aligned}
\end{equation}
Here, $\mathcal{D}$ represents any distance to measure the distance between the user/item and its cluster center, and $\rho$ defines the maximum allowed proximity between cluster centers.
Additionally, to ensure that users and items are close to their assigned cluster centers, we compute the mean squared error (MSE) between the latent vectors and their corresponding cluster centers:

\begin{equation}\label{equ:meanu}
\begin{aligned}
    loss_{1,c}^{user}=\frac{1}{N}\sum_{j=0}^{v} \sum_{i=0}^{t}\sum_{k=0}^{N,k \in S_{i,j}}\Vert c_{i,j}^{user}-p_k^{user}\Vert^2 \\
    loss_{1,c}^{item}=\frac{1}{N}\sum_{j=0}^{v} \sum_{i=0}^{t}\sum_{k=0}^{N,k \in S_{i,j}}\Vert c_{i,j}^{item}-q_k^{item}\Vert^2 \\
\end{aligned}
\end{equation}
where $S_{i,j}$ is the set of user/item that belongs to the $i^{th}$ cluster in the $j^{th}$ view. To make full use of the representation space, the objective to optimize for cluster centers can be defined as:
\begin{equation}
    \label{equ:loss1a}
    loss1 = loss_1^{user}+loss_1^{item}+loss_{1,c}^{user}+loss_{1,c}^{item}
\end{equation}

Considering the difficulty of determining the number of cluster centers that accurately represent potential space, it is also challenging to ascertain if there is redundancy in our problem. 
To address this problem, we introduce a pruning mechanism that removes redundant cluster centers during training to dynamically manage the number of cluster centers. After each iteration $I_p$, we compute the mean weight of each cluster center and remove those with weights below a threshold $\psi$. This process is outlined in Algorithm \ref{alg:alg1}.
\begin{algorithm}[t!]
\small
\caption{Dynamic clustering algorithm}
\label{alg:alg1}
\textbf{Input}: Cluster centers: $\mathcal{C}$; Weights of user/item: $\mathcal{W}$; Epoch: $i$.\\
\textbf{Output}: New cluster centers: $\mathcal{C}'$; Clustering loss: $\mathcal{L}_{1,c}$
\begin{algorithmic}[1] 
\IF {$i > I_p$}
\STATE $\overline{\mathcal{W}} \leftarrow $ cluster-wise mean of user/item in $W$.
\STATE $\mathcal{C}',\mathcal{W}\leftarrow $ Remove the $\mathcal{C}$ and $\mathcal{W}$ where $\overline{\mathcal{W}} < \psi$.
\ELSE
\STATE $\mathcal{C}' \leftarrow \mathcal{C}$.
\ENDIF
\STATE $\mathcal{P},\mathcal{Q} \leftarrow \mathcal{W}$ weighted sum $\mathcal{C}'$.
\STATE $\mathcal{L}_{1,c} \leftarrow$ Compute using $\mathcal{C}'$, $\mathcal{P}$, $\mathcal{Q}$ and Eq.\ref{equ:meanu}.
\STATE \textbf{return} $\mathcal{C}'$, $\mathcal{L}_{1,c}$
\end{algorithmic}
\end{algorithm}
\begin{figure}[b!]
   \includegraphics[width=\columnwidth]{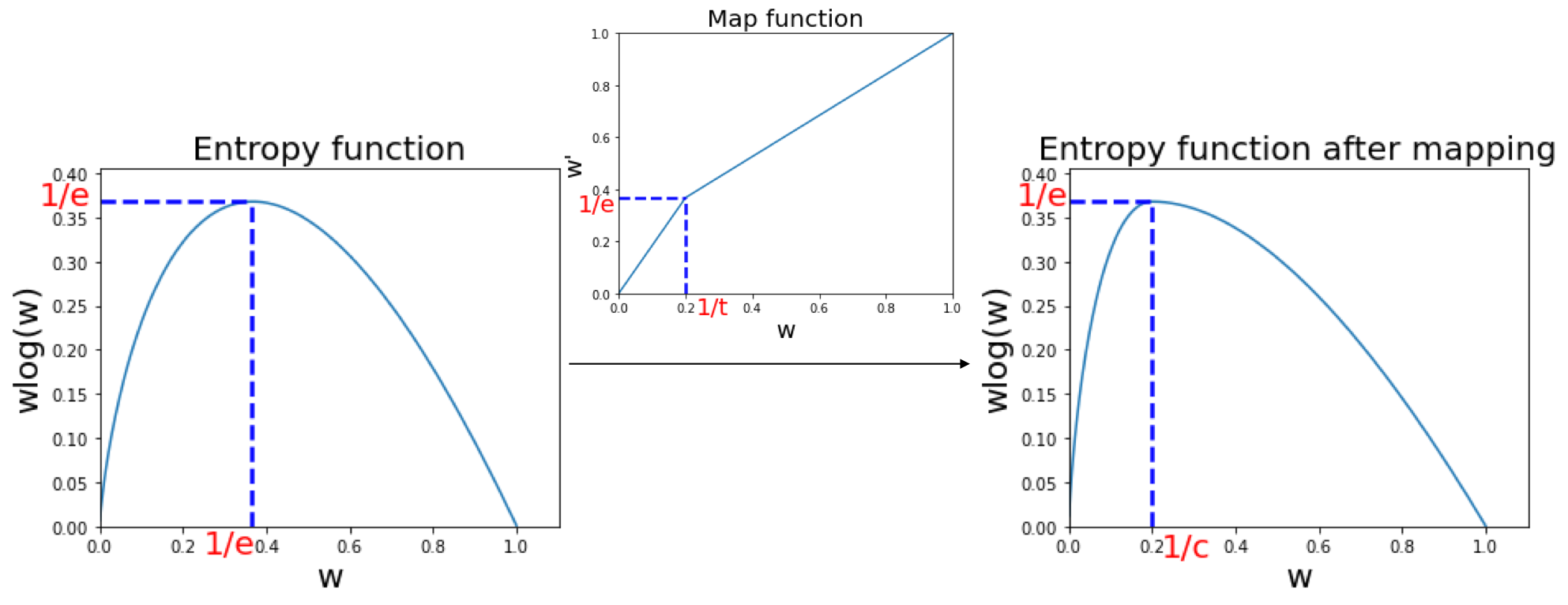}
    \caption{Mapping function of Eq. \ref{equ:map} for the weight. Using the function in the middle of the Figure, we can map the weights on the left to the right, which facilitates a relatively balanced optimization of weight in different views.
    }\label{fig:map}
\end{figure}

\subsection{User/item weights}  
To ensure interpretability and optimization across views, we normalize the user/item weights using the softmax function:
\begin{equation}
\small
\begin{aligned}
    \label{equ:loss21}
    W^{user'} = \operatorname{softmax}(W^{user}) ,
    W^{item'} = \operatorname{softmax}(W^{item}) 
\end{aligned}
\end{equation}

Additionally, the number of centers varies from view to view. In views with a different number of centers, the losses behave differently even if the weight distribution is the same. For example, in $view_1$ and $view_2$, there are 3 and 10 centers respectively, and the weight distribution of users is uniform. Although the distribution of both views is poor, the loss of $view_1$ is almost at the extreme point and much greater than the loss of $view_2$. This can lead to unbalanced optimizations, meaning that it is much more difficult to optimize losses in views with more centers. To this end, we introduce a mapping function to address the issue of unbalanced optimization across views with varying numbers of cluster centers. The mapping function ensures that the uniform distribution loss (worst-case scenario) remains consistent across views, regardless of the number of centers. The mapping function is defined as:
\begin{equation}
    \label{equ:map}
    w''=\left\{\begin{array}{ccl} 
        \frac{t}{e}w' && 0 \leq w' \leq \frac{1}{t} \\
        \frac{tw'-1}{t-1}(1-\frac{1}{e})+\frac{1}{e} && \frac{1}{t}\le w' \le 1 
    \end{array} \right.
\end{equation}
Here, $t$ is the number of cluster centers in a view, and $e$ is the base of the natural logarithm. The entropy-based loss for user/item weights is then computed as:
\begin{equation}
\label{equ:loss23}
\begin{aligned}
    loss_2^{user} = -\sum_{j=0}^{v} \sum_{i=0}^{m} w_{i,j}^{user''}log(w_{i,j}^{user''}) \\
    loss_2^{item} = -\sum_{j=0}^{v} \sum_{i=0}^{n} w_{i,j}^{item''}log(w_{i,j}^{item''}) 
\end{aligned}
\end{equation}

The total loss for user/item weights is:
\begin{equation}
\label{equ:loss2}
    loss_2 = loss_2^{item} + loss_2^{user}
\end{equation}

\subsection{Total Objective Function}
The overall objective function of MFDMC combines the losses for cluster centers, user/item weights, and the target user-item interaction ratings:
\begin{equation}
    \label{equ:loss}
    loss = \eta loss_1 + \gamma loss_2 + loss_3
\end{equation}
Here, $\eta$ and $\gamma$ are hyperparameters that control the relative importance of the cluster center and weight losses, respectively.
The term $loss_3$ represents the loss associated with the user-item interaction matrix, which is typically computed using a matrix factorization objective (e.g., mean squared error or cross-entropy).
\subsection{Dynamic Multi-view Clustering for Adaptive Downstream Tasks in Computer Vision}
\begin{figure}[t!]
       \includegraphics[width=0.95\columnwidth]{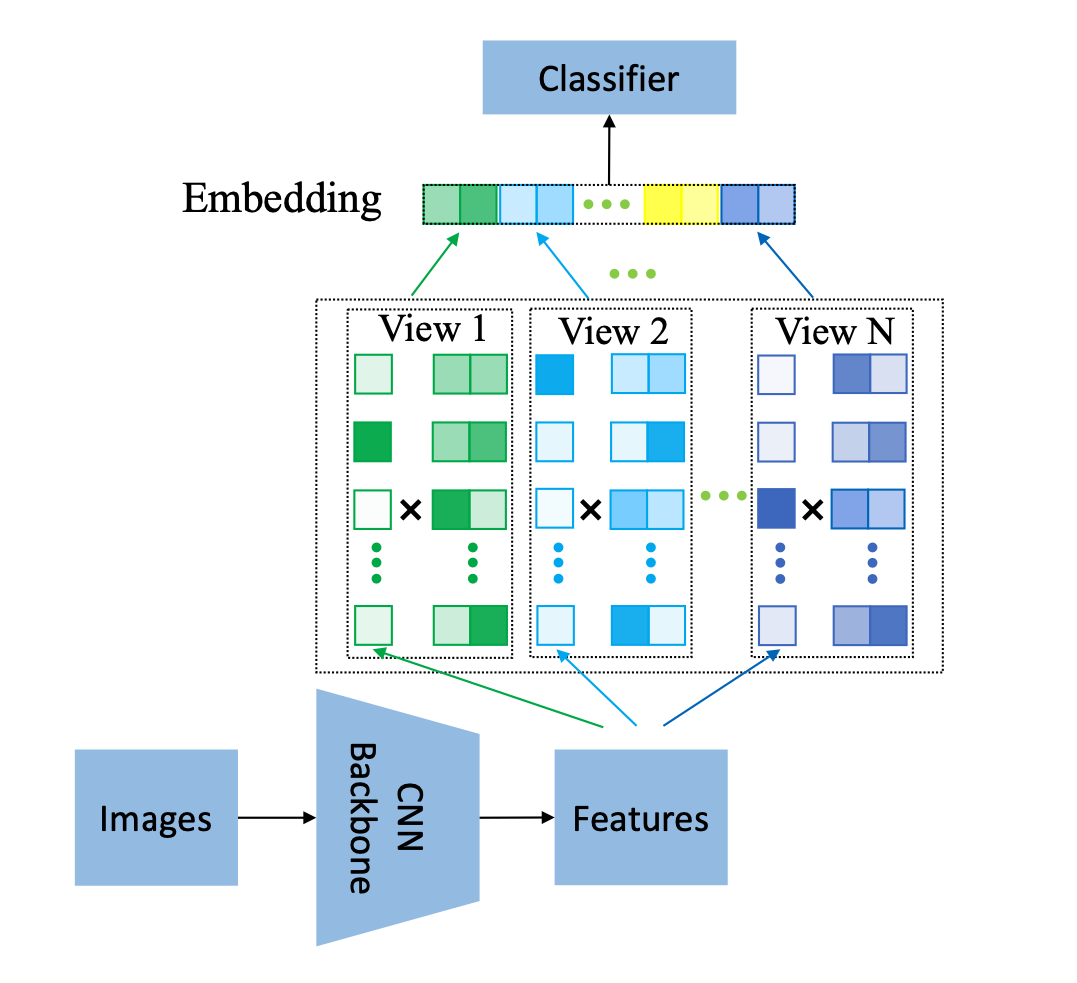}
        \caption{Illustration of the MFDMC for computer vision tasks. The features are embedded into the weights of cluster centers in each view. The final representation of each image for downstream tasks is obtained by summing up the weighted centers of each view and concatenating them.
        }\label{fig:cv_method}
        \vspace*{-0.2in}
    \end{figure}

The framework of Dynamic Multi-view Clustering (DMC) in Computer Vision (CV) is depicted in Fig.~\ref{fig:cv_method}. This approach requires minimal modifications to existing architectures. Specifically, features extracted via CNN are embedded into the weights of cluster centers for each view. The final representation of each image, intended for downstream tasks, is derived by aggregating the weighted cluster centers across all views and concatenating them.
The loss function for optimizing the cluster centers is formulated in Eq.~\ref{eq:16}, which comprises two components: first a dispersion loss that encourages cluster centers to be well-separated, as defined in Eq.~\ref{eq:14}, and second a proximity loss that ensures images are closely aligned with their respective cluster centers, as specified in Eq.~\ref{eq:15}. Additionally, a regularization term, presented in Eq.~\ref{eq:18}, is introduced to stabilize the cluster weights. Finally, the cross-entropy loss for classification is expressed in Eq.~\ref{eq:19}, where $N_c$ denotes the number of image classes.
The framework is designed to demonstrate the scalability and versatility of our MFDMC , thereby improving performance in downstream tasks such as image classification.

\begin{equation}
    \text{loss}_1^{cv} = \sum_{j=1}^{v} \sum_{\alpha,\beta} l(c_{\alpha,j}^{cv}, c_{\beta,j}^{cv})
    \label{eq:14}
\end{equation}

\begin{equation}
    l(c_{\alpha}, c_{\beta}) = \max \{ 0, \rho - \mathcal{D}(c_{\alpha} - c_{\beta}) \}
    \notag
\end{equation}

\begin{equation}
    \text{loss}_{1,c}^{cv} = \frac{1}{|S_{i,j}|} \sum_{j=1}^{v} \sum_{i=1}^{t} \sum_{k=1}^{|S_{i,j}|} \| c_{i,j}^{cv} - p_k^{cv} \|^2
    \label{eq:15}
\end{equation}

\begin{equation}
    \text{loss}_1 = \text{loss}_1^{cv} + \text{loss}_{1,c}^{cv}
    \label{eq:16}
\end{equation}

\begin{equation}
    W^{cv'} = \text{softmax}(W^{cv})
    \label{eq:17}
\end{equation}

\begin{equation}
    \text{Loss}_2 = -\sum_{j=1}^{v} \sum_{i=1}^{t_j} w'_{i,j} \log \left( \frac{w'_{i,j}}{t_j} \right)
    \label{eq:18}
\end{equation}

\begin{equation}
    \text{loss}_3 = -\frac{1}{N} \sum_{i=1}^{N} \sum_{c=1}^{N_c} y_{ic} \log (\hat{y}_{ic})
    \label{eq:19}
\end{equation}

\section{Experiments}
\label{sec:experiments}
In this section, we present extensive experimental results on six real-world datasets to validate the effectiveness of our proposed approach.
To demonstrate the superiority of our MFDMC model, we conduct comprehensive comparisons with traditional matrix factorization methods and state-of-the-art techniques. 
These comparisons highlight the enhanced performance and interpretability of our model.
Furthermore, leveraging insights from the interpretability analysis in Section~\ref{sec5:interpre}, we extend our evaluation to include a detailed examination of cluster semantics and user/item representations. 

\subsection{Datasets \& Implementation} 
\begin{table}[t!]
\centering
\footnotesize
\caption{Statistics of the datasets.
}
\begin{tabular}{p{2.2cm}|p{0.9cm}p{0.85cm}p{1.05cm}p{1.3cm}}
    \hline
    Dataset & $m$ & $n$ & $N$ & Range\\
    \hline
    MovieLens-1M & 6,040 & 3,952 & 1,000,209 & [1, 5] \\ 
    MovieLens-100k & 943 & 1,682 & 100,000 & [1, 5]  \\
    Amazon-video & 424,560 & 23,745 & 583,933 & [1, 5]  \\ 
    Epinions & 40,163 & 139,738 & 664,824 & [1, 5]  \\ 
    Books-across & 105,283 & 340,395 & 1,149,780 &[0, 10] \\ 
    Jester & 73,421 & 100 &	3,519,446 & [-10, 10]\\ 
    \hline
\end{tabular}
\label{dataset}
\end{table}
\textit{Datasets for recommender system.}
We evaluate our proposed MFDMC on six real-world benchmarks, i.e.,
\textit{MovieLen} \cite{dataset:movielens} , \textit{Amazon-video}~\cite{dataset:amazon}, \textit{Epinions}\footnote{\url{http://www.trustlet.org/datasets/downloaded_epinions/}}, \textit{Books-across}~\cite{dataset:books}, and the \textit{Jester} dataset \cite{dataset:jester}.
Detailed statistics are summarized in Table \ref{dataset}.

\textit{Datasets for computer vision} To validate the interpretability of our method in the context of computer vision, we construct a synthetic toy dataset. This dataset consists of 9 distinct classes of images, categorized by shape and color: green triangles, green rectangles, green circles, red triangles, red rectangles, red circles, blue triangles, blue rectangles, and blue circles. Each class contains 200 images, resulting in a balanced dataset for evaluation. This dataset is designed to test the ability of our method to distinguish and interpret visual features effectively.
\subsection{Implementation details.} 
For each dataset, the data is partitioned into training, validation, and testing sets. Specifically, 80\% of the data is randomly selected for training the model, while the remaining 20\% is equally divided into validation and testing sets (10\% each). This ensures a robust evaluation of the model's performance.
To optimize the hyperparameters, we employ the \textit{Optuna} framework \cite{akiba2019optuna}, a state-of-the-art automatic hyperparameter tuning tool. Specifically, $I_d$ is set to 40, which means that in the first 40 epochs, the cluster centers are not removed dynamically in all experiment.
The threshold to remove the cluster center is $\frac{1}{t}$, and we keep at least 3 centers in one view.
The weight parameters $\eta$ and $\gamma$ of $loss_1$ and $loss_2$ gradually increase with epochs.
As the number of views for users and items is the same, centers in the same location in the view can be shared.
And the number of cluster centers $t$ in each view is 10.
Weight decay with a regularization parameter $\lambda$ is also added to our model.
Distance $\mathcal{D}$ in Eq. \ref{equ:loss1u} is used to measure the distance between user/item and its cluster center.
The Euclidean distance is used as a metric, calculated as:
\begin{equation}
    \label{equ:loss1d}
    \mathcal{D}(c_{\alpha}, c_{\beta}) = \Vert c_{\alpha} - c_{\beta} \Vert^2
\end{equation}
With cluster centers are view-wise normalized by:
\begin{equation}
    \label{equ:norm}
    C = \frac{C-min(C)}{max(C)-min(C)}
\end{equation}

\textbf{Evaluation metric} Root mean square error (RMSE) is used as an evaluation metric, formulated as
\begin{equation}
RSME = \sqrt{\frac{1}{N}\sum_{i=0}^{N}(y_i - r_i)^2}.
\end{equation}
\subsection{Results \& Analysis}
\begin{table}[t]
\centering
\footnotesize
\caption{Comprehensive comparison results
across the MovieLens-1M and MovieLens-100k datasets for various baselines. The token (D) specifies the dimension of the latent vector $d$.}
\resizebox{.9\linewidth}{!}{
\begin{tabular}{lll}
    \hline
    Method & MovieLens-1M & MovieLens-100k \\
    \hline
    FunkMF(16) &  0.869  & 0.938\\
    FunkMF(60) & 0.868  & 0.936 \\
    BiasedMF(16) & 0.866 & 0.926\\
    BiasedMF(60) & 0.863 & 0.923 \\
    PMF(60) & 0.883 & 0.952\\
    \hline
    SVD & 0.884 & 0.942 \\
    Glocal-K & 0.863 & 0.923 \\
   \hline
    Ablation(16) & 0.862  & 0.943 \\
    MFDMC(16) & \textbf{0.848} & \textbf{0.911}\\
    \hline
\end{tabular}}
\label{table:movie_result}
\end{table}

\begin{table}[t]
\centering
\footnotesize
\caption{Comprehensive comparison results
across the four datasets for various baselines. The token (D) specifies the dimension of the latent vector $d$.}
\resizebox{.9\linewidth}{!}{
\begin{tabular}{lllll}
    \hline
    Method & Amazon-video & Epinions & Books-across & Jester\\
    \hline
    FunkMF(16) & 2.063 & 1.637& 3.836 & 4.973\\
    FunkMF(60)  & 2.163 & 1.776 & 3.863 & 5.049\\
    BiasedMF(16) & 1.034 & 1.039 & 3.508 & 4.904\\
    BiasedMF(60) & 1.030 & 1.044 & 3.503 & 4.906\\
   \hline
    Ablation(16)  &1.036 & 1.043 & 3.398 & 4.220\\
    MFDMC(16) & \textbf{1.019} & \textbf{1.034} &  \textbf{3.352} & \textbf{4.114}\\
    \hline
\end{tabular}}
\label{table:result}
\end{table}
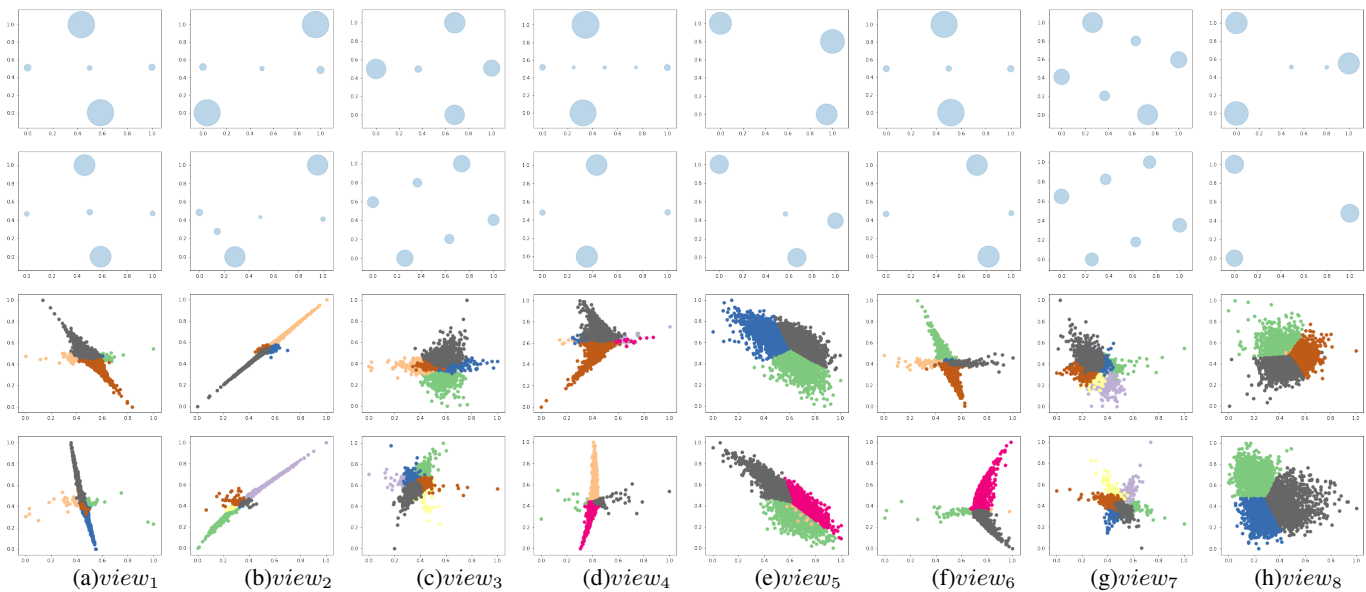
\begin{figure*}[t!]
  \includegraphics[width=\linewidth]{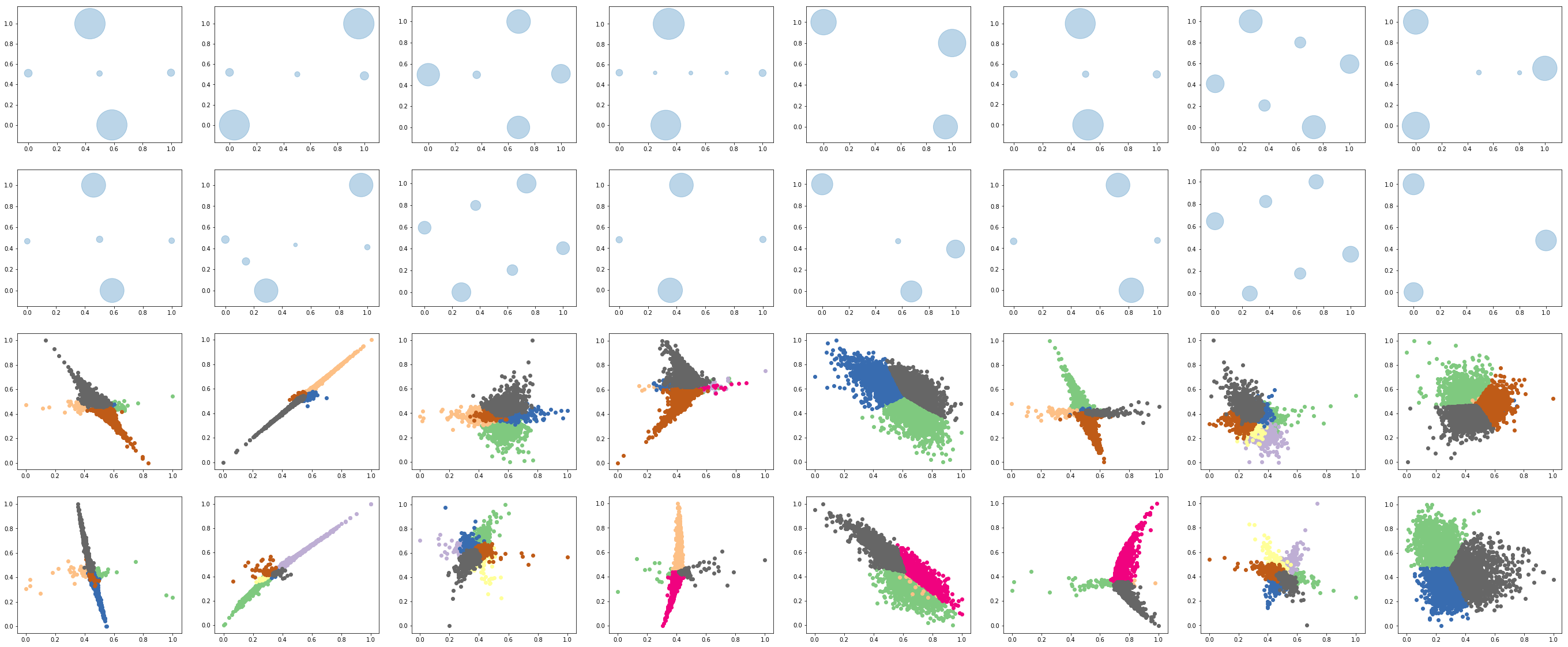}\small
    \put(-490,-7){(a)$view_1$}
    \put(-425,-7){(b)$view_2$}
    \put(-360,-7){(c)$view_3$}
    \put(-297,-7){(d)$view_4$}
    \put(-232,-7){(e)$view_5$}
    \put(-165,-7){(f)$view_6$}
    \put(-105,-7){(g)$view_7$}
    \put(-43,-7){(h)$view_8$}
  \caption{Multi-view Clustering results. The four rows show the clustering centers of users/items, and the clustering results of users/items respectively. Each column represents a view. In the clustering result, each point corresponds to the value of the user/item latent vector in a specific view. }\label{fig:cluster}
\end{figure*} Table~\ref{table:movie_result} and ~\ref{table:result} report the RMSE values for different methods across real-world datasets.
Concretely, MFDMC consistently outperforms the other competitors, with an improvement of approximately 0.025 RMSE compared to FunkMF on the MovieLens-100k. Despite the increase in the dimension of the latent space from 16 to 60 in FunkMF\cite{result} and BiasedMF~\cite{biased_mf}, the RMSE did not improve or worsen. This suggests that these models are unable to fully utilize the representation space.
In contrast, MFDMC, with a latent space dimension of 16 (same as FunkMF), achieves significantly better RMSE results. Interestingly, it is observed that even with a lower dimension of 12 (Table~\ref{table:config}), MFDMC can achieve results comparable to those of other methods, such as SVD~\cite{jadon2023comprehensive}, PMF~\cite{pmf} and Glocal-K\cite{han2021glocal}.
Furthermore, the effectiveness of our approach is attributed not only to the deep structure of our model but also to the loss functions we designed. Without these loss functions, the RMSE of our model is only slightly better than or even worse than the other methods. However, with the inclusion of loss functions, the RMSE improves consistently.

\begin{figure*}[t]
    \centering
    \includegraphics[width=0.8\linewidth]{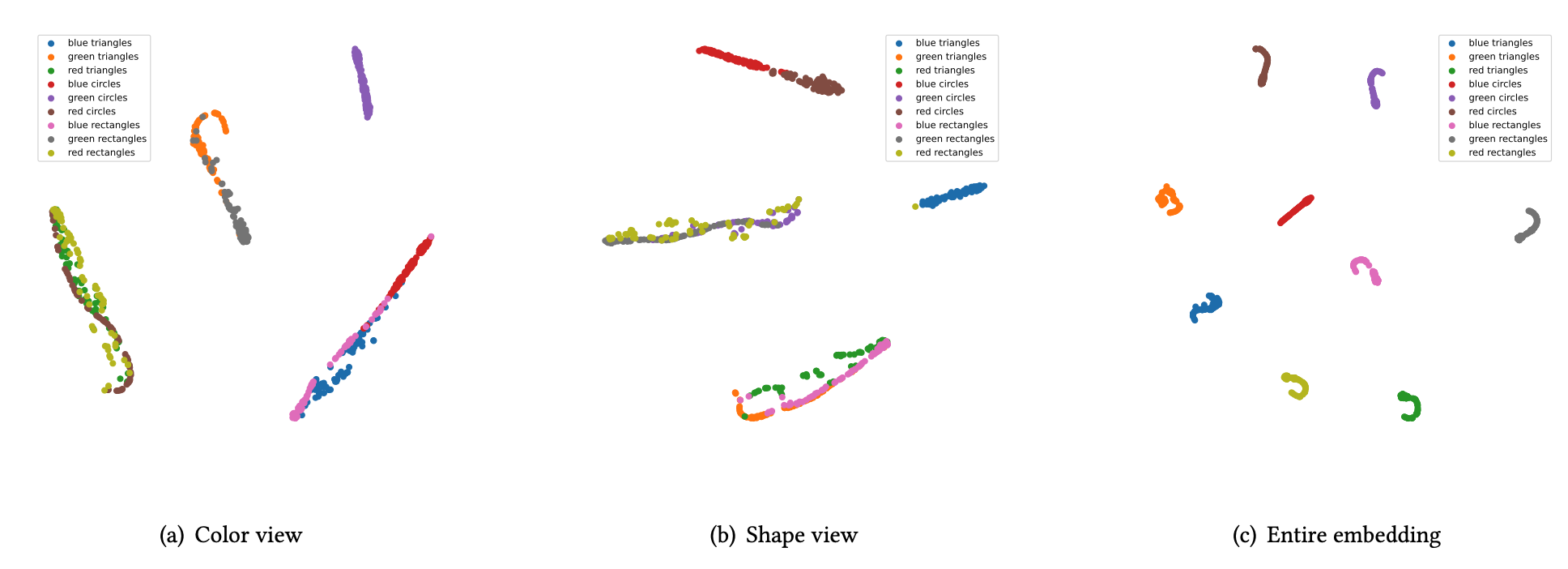}
    \caption{The t-sne results of experiments on synthetic toy dataset. (a), (b), and (c) show the results of shape, color view, and the entire embedding respectively.}
    \label{fig:cv}
\end{figure*}

In addition, extended experiments on down-stream computer vision tasks are conducted to further verify the interpretability and adaptivity of our method.
The results are shown in Fig.~\ref{fig:cv}. In the color view of Fig.~\ref{fig:cv}(a), the green, red and blue geometric shapes are grouped together respectively. In the shape view of Fig.~\ref{fig:cv}(b), the green rectangles and red rectangles, blue circles and red circles, green triangles and red triangles are clustered together correctly. Fig.~\ref{fig:cv}(c) shows the t-sne results of the entire embedding, the 9 classes of shapes are correctly clustered.

\subsection{Ablation Study}  
\label{sec:ablation}
We conducted ablation experiments on the MovieLens-1M dataset to evaluate the impact of various configurations on RMSE. Increasing the number of views improved RMSE initially, but performance degraded with excessive views. Sharing centers between users and items had negligible effects, while increasing the latent space dimension did not significantly enhance results. Based on these findings, we recommend selecting the number of views based on the desired class granularity, sharing centers to optimize resource usage, and maintaining a low-dimensional latent space for efficient performance.
\begin{table}[h]
\centering
\small
\footnotesize
\caption{Ablation study of different configurations.}
\resizebox{.85\columnwidth}{!}{
\begin{tabular}{llllll}
    \hline
    No. & Not Share & Share & $v$  & $d$ & $b$\\
    \hline
    1 & 0.862 & 0.863 &2 & 12 & 6 \\
    2 & 0.857 & 0.854 & 4 & 12 & 3 \\
    3 & 0.856 & 0.852 & 6 & 12 & 2 \\
    4 & 0.849 & 0.850 & 8 & 16 & 2 \\
    5 & 0.851 & 0.852 & 8 & 24 & 3 \\
    6 & 0.851 & 0.851 & 10 & 20 & 2 \\
    \hline
\end{tabular}}
\label{table:config}
\end{table}
\begin{figure}[t] 
\centering
\includegraphics[width=0.9\columnwidth]{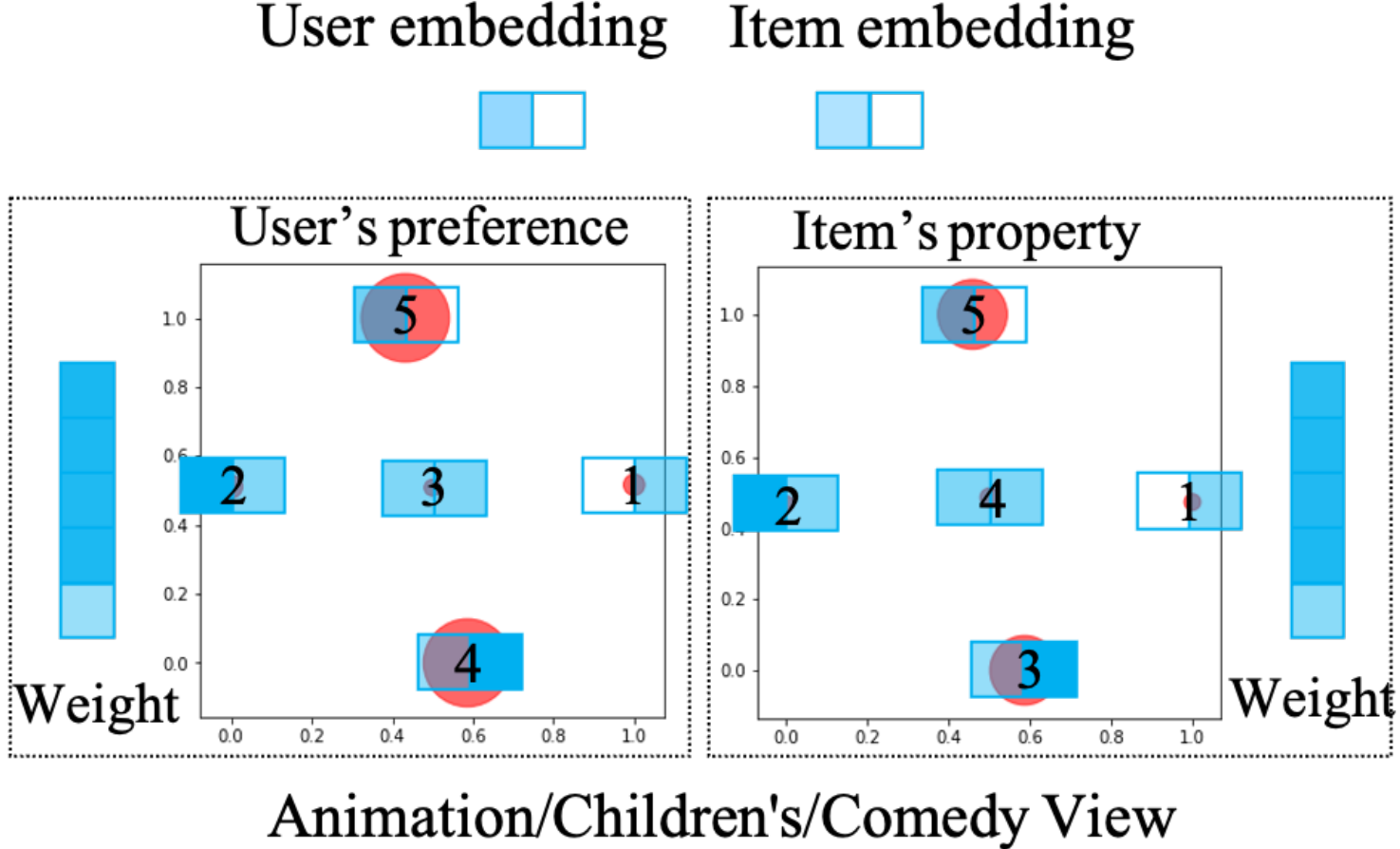}
    \caption{Illustration of interpreting a representation.
    Red circles depict the cluster's center, while squares represent the embedding of the centers and weights. The transparency of each shape reflects its value, and the number in the center indicates the position of its corresponding weights.
    }\label{fig:illustration}
\end{figure}
\subsection{Interpretability Analysis of User/Item Representations}  
\label{sec5:interpre}
To demonstrate the interpretability of the proposed method, we analyze user/item representations using an instance from Experiment No. 4 (Table~\ref{table:config}). Visualizing movie distributions across clusters reveals that animation, children, and comedy movies span multiple clusters, while other categories concentrate in single clusters, suggesting $view_1$ represents these genres. To further interpret user preferences, we analyze rating distributions within clusters (Fig.~\ref{fig:illustration}). Clusters with higher average ratings for specific categories, such as animation, indicate stronger user preferences or perceived quality for those genres. This analysis aligns with Explainable AI (XAI) principles \cite{rotem2024visuala,rotem2024visualb}, linking clusters to semantic attributes like movie categories and user preferences. By providing transparent insights into cluster meanings, our method enhances model trustworthiness and supports actionable, personalized recommendations.

\section*{Acknowledgments}
This research was partially supported by Zhejiang Key R\&D Program of China under grant No. 2024C03048, Zhejiang Key Laboratory of Medical Imaging Artificial Intelligence, State Key Laboratory of Transvascular Implantation Devices under Grant NO. SKLTID2024003, and the Transvascular lmplantation Devices Research Institute (TIDRI).
\section*{Conclusion}
In this study, we propose a novel approach, called Unified Matrix Factorization with Dynamic Multi-view Clustering (MFDMC). Our motivation stems from the observation that the representation space is not fully utilized in traditional matrix factorization (MF) algorithms, leading to uninterpretable user/item representations. Additionally, downstream clustering tasks require significant additional time and resources. To address these issues, we introduce dynamic multi-view clustering to MF. By incorporating dynamic multi-view clustering into MF, our method not only enhances the interpretability of representations but also optimally utilizes the representation space. Extensive experiments on massive datasets demonstrate the superiority of our proposed MFDMC over existing MF methods. In future work, we plan to further improve our approach and extend its applicability to a wider range of models for downstream tasks.

\bibliographystyle{IEEEtran}
\bibliography{IJCNN25_DMC}
\end{document}